\documentclass[12pt,a4paper]{article}
\makeatletter
\def\eqnarray{%
\stepcounter{equation}%
\let\@currentlabel=\theequation
\global\@eqnswtrue
\global\@eqcnt\z@
\tabskip\@centering
\let\\=\@eqncr
$$\halign to \displaywidth\bgroup\@eqnsel\hskip\@centering
$\displaystyle\tabskip\z@{##}$&\global\@eqcnt\@ne
\hfil$\displaystyle{{}##{}}$\hfil
&\global\@eqcnt\tw@$\displaystyle\tabskip\z@{##}$\hfil
\tabskip\@centering&\llap{##}\tabskip\z@\cr}
\makeatother
\newcommand{\kansu}[2]{{{#1}\!\left({#2}\right)}}
\newcommand{\ket}[1]{{\vert{#1}\rangle}}
\newcommand{\bra}[1]{{\langle{#1}\vert}}
\newcommand{\kett}[2]{{\vert{#1,#2}\rangle}}
\newcommand{\calh}{{\cal H}}
\newcommand{\calm}{{\cal M}}
\newcommand{\cala}{{\cal A}}
\newcommand{\calf}{{\cal F}}

\newcommand{\fukuso}{{\mathbf C}}

\newcommand{\futon}{{\bf N}}

\newcommand{\stm}{{St_m}}
\newcommand{\grm}{{Gr_m}}
\newcommand{\eem}{{E_m}}

\newcommand{\lam}{{\bf \lambda}}
\newcommand{\slam}{{\bf \lambda_0}}

\begin{document}

\title{\sl From Geometry to Quantum Computation
           \thanks{A talk at the 2nd International Symposium ``
           Quantum Theory and Symmetries'', Krakow, 18--21, 
           July, 2001}
      }
\author{
  Kazuyuki FUJII
  \thanks{E-mail address : fujii@math.yokohama-cu.ac.jp }\\
  Department of Mathematical Sciences\\
  Yokohama City University\\
  Yokohama 236-0027\\
  JAPAN
  }
\date{}
\maketitle\thispagestyle{empty}
%
%
%  gaiyou
%
%
\begin{abstract}
  The aim of this paper is to introduce our idea of Holonomic Quantum 
  Computation (Computer). Our model is based on both harmonic oscillators 
  and non--linear quantum optics, not on spins of usual quantum computation 
  and our method is moreover completely geometrical. 

  We hope that therefore our model may be strong for decoherence.
\end{abstract}

\newpage

%
%
%     Honbun
%
%

\section{Introduction}

Quantum Computation is a very attractive and challenging task in 
this century.
\par \noindent
After the breakthrough by P. Shor \cite{PS} there has been remarkable
progress in Quantum Computer or Computation (QC briefly).
This discovery had a great influence on scientists. This drived not only 
theoreticians to finding other quantum algorithms, but also 
experimentalists to building quantum computers.
See \cite{AS} and \cite{RP}, \cite{KF9} in outline. 

On the other hand, Gauge Theories are widely recognized as the basis in 
quantum field theories.
Therefore it is very natural to intend to include gauge theories 
in QC $\cdots$ a construction of ``gauge theoretical'' quantum computation
or of ``geometric'' quantum computation in our terminology. 
The merit of geometric method of QC may be strong for the influence 
from the environment.  

In \cite{ZR} and \cite{PZR} Zanardi and Rasetti proposed an attractive 
idea $\cdots$ Holonomic Quantum Computation (Computer) $\cdots$
using the non-abelian Berry phase (quantum holonomy in the mathematical 
language). 
See also \cite{AYK} and \cite{JP} as another interesting geometric models.

In their model a Hamiltonian (including some parameters) must be
degenerated because an adiabatic connection is introduced using
this degeneracy \cite{SW}. 
In other words, a quantum computational bundle is introduced on some 
parameter space due to this degeneracy (see \cite{ZR}) and the canonical 
connection of this bundle is just the above.

They gave a few simple but interesting examples to explain their idea. 
To make their works more mathematical and rigorous the author has 
given the mathematical reinforcement to their works, 
see \cite{KF1}, \cite{KF3}, \cite{KF4} and \cite{KF5}. 
But his works are still not sufficient. 

In this talk we will introduce our Holonomic Quantum Computation 
and discuss some problems to be solved. 

We strongly hope that young mathematical physicists will enter this 
attractive field.

\section{Mathematical Preliminaries}

We start with mathematical preliminaries.
Let $\calh$ be a separable Hilbert space over $\fukuso$.
For $m\in{\bf N}$, we set
\begin{equation}
  \label{eq:stmh}
  \kansu{\stm}{\cal H}
  \equiv
  \left\{
    V=\left( v_1,\cdots,v_m\right)
    \in
    \calh\times\cdots\times\calh\vert V^\dagger V=1_m\right\}\ ,
\end{equation}
($V^\dagger V=1_m \Longleftrightarrow 
\langle v_{i}\vert v_{j}\rangle = \delta_{ij}$) where $1_m$ is a unit matrix 
in $\kansu{M}{m,\fukuso}$.
This is called a (universal) Stiefel manifold.
Note that the unitary group $U(m)$ acts on $\kansu{\stm}{\calh}$
from the right:
\begin{equation}
  \label{eq:stmsha}
  \kansu{\stm}{\calh}\times\kansu{U}{m}
  \rightarrow
  \kansu{\stm}{\calh}\  :\  \left( V,a\right)\mapsto Va.
\end{equation}
Next we define a (universal) Grassmann manifold
\begin{equation}
  \kansu{\grm}{\calh}
  \equiv
  \left\{
    X\in\kansu{M}{\calh}\vert
    X^2=X, X^\dagger=X\  \mathrm{and}\  \mathrm{tr}X=m\right\}\ ,
\end{equation}
where $M(\calh)$ denotes a space of all bounded linear operators on $\calh$.
Then we have a projection
\begin{equation}
  \label{eq:piteigi}
  \pi : \kansu{\stm}{\calh}\rightarrow\kansu{\grm}{\calh}\ ,
  \quad \kansu{\pi}{V}\equiv VV^\dagger\ ,
\end{equation}
compatible with the action (\ref{eq:stmsha}) 
($\kansu{\pi}{Va}=Va(Va)^\dagger=Vaa^\dagger V^\dagger=VV^\dagger=
\kansu{\pi}{V}$).

Now the set
\begin{equation}
  \label{eq:principal}
  \left\{
    \kansu{U}{m}, \kansu{\stm}{\calh}, \pi, \kansu{\grm}{\calh}
  \right\}\ ,
\end{equation}
is called a (universal) principal $U(m)$ bundle, 
see \cite{MN} and \cite{KF1}.\quad We set
\begin{equation}
  \label{eq:emh}
  \kansu{\eem}{\cal H}
  \equiv
  \left\{
    \left(X,v\right)
    \in
    \kansu{\grm}{\calh}\times\calh \vert Xv=v \right\}\ .
\end{equation}
Then we have also a projection 
\begin{equation}
  \label{eq:piemgrm}
  \pi : \kansu{\eem}{\calh}\rightarrow\kansu{\grm}{\calh}\ ,
  \quad \kansu{\pi}{\left(X,v\right)}\equiv X\ .
\end{equation}
The set
\begin{equation}
  \label{eq:universal}
  \left\{
    \fukuso^m, \kansu{\eem}{\calh}, \pi, \kansu{\grm}{\calh}
  \right\}\ ,
\end{equation}
is called a (universal) $m$-th vector bundle. This vector bundle is 
one associated with the principal $U(m)$ bundle (\ref{eq:principal})
.

Next let $M$ be a finite or infinite dimensional differentiable manifold 
and the map $P : M \rightarrow \kansu{\grm}{\calh}$ be given (called a 
projector). Using this $P$ we can make 
the bundles (\ref{eq:principal}) and (\ref{eq:universal}) pullback 
over $M$ :
\begin{eqnarray}
  \label{eq:hikimodoshi1}
  &&\left\{\kansu{U}{m},\widetilde{St}, \pi_{\widetilde{St}}, M\right\}
  \equiv
  P^*\left\{\kansu{U}{m}, \kansu{\stm}{\calh}, \pi, 
  \kansu{\grm}{\calh}\right\}
  \ , \\
  \label{eq:hikimodoshi2}
  &&\left\{\fukuso^m,\widetilde{E}, \pi_{\widetilde{E}}, M\right\}
  \equiv
  P^*\left\{\fukuso^m, \kansu{\eem}{\calh}, \pi, \kansu{\grm}{\calh}\right\}
  \ ,
\end{eqnarray}
see \cite{MN}. (\ref{eq:hikimodoshi2}) is of course a vector bundle 
associated with (\ref{eq:hikimodoshi1}).

Let $\calm$ be a parameter space and we denote by $\lam$ its element. 
Let $\slam$ be a fixed reference point of $\calm$. Let $H_\lam$ be 
a family of Hamiltonians parameterized by $\calm$ which act on a Fock space 
$\calh$. We set $H_0$ = $H_\slam$ for simplicity and assume that this has 
a $m$-fold degenerate vacuum :
\begin{equation}
  H_{0}v_j = \mathbf{0},\quad j = 1 \sim m. 
\end{equation}
These $v_j$'s form a $m$-dimensional vector space. We may assume that 
$\langle v_{i}\vert v_{j}\rangle = \delta_{ij}$. Then $\left(v_1,\cdots,v_m
\right) \in \kansu{\stm}{\calh}$ and 
\[
  F_0 \equiv \left\{\sum_{j=1}^{m}x_{j}v_{j}\vert x_j \in \fukuso \right\} 
  \cong \fukuso^m.
\]
Namely, $F_0$ is a vector space associated with o.n.basis 
$\left(v_1,\cdots,v_m\right)$.

Next we assume for simplicity 
that a family of unitary operators parameterized by $\calm$
\begin{equation}
  \label{eq:ufamily} 
  W : \calm \rightarrow U(\calh),\quad W(\slam) = {\rm id}.
\end{equation}
is given and $H_{\lam}$ above is given by the following isospectral family
\begin{equation}
 H_{\lam} \equiv W(\lam)H_0 W(\lam)^{-1}.
\end{equation}
In this case there is no level crossing of eigenvalues. Making use of 
$W(\lam)$ we can define a projector
\begin{equation}
  \label{eq:pfamily}
 P : \calm \rightarrow \kansu{\grm}{\calh}, \quad 
 P(\lam) \equiv
  W(\lam) \left(\sum^{m}_{j=1}v_{j}v_{j}^{\dagger}\right)W(\lam)^{-1}
\end{equation}
and have the pullback bundles over $\calm$
\begin{equation}
  \label{eq:target}
 \left\{\kansu{U}{m},\widetilde{St}, \pi_{\widetilde{St}}, \calm\right\},\quad 
 \left\{\fukuso^m,\widetilde{E}, \pi_{\widetilde{E}}, \calm\right\}.
\end{equation}

For the latter we set
\begin{equation}
  \label{eq:vacuum}
 \ket{vac} = \left(v_1,\cdots,v_m\right).
\end{equation}
In this case a canonical connection form $\cala$ of 
$\left\{\kansu{U}{m},\widetilde{St}, \pi_{\widetilde{St}}, \calm\right\}$ is 
given by 
\begin{equation}
  \label{eq:cform}
 \cala = \bra{vac}W(\lam)^{-1}d W(\lam)\ket{vac},
\end{equation}
where $d$ is a differential form on $\calm$, and its curvature form by
\begin{equation}
  \label{eq:curvature}
  \calf \equiv d\cala+\cala\wedge\cala,
\end{equation}
see \cite{SW} and \cite{MN}.

Let $\gamma$ be a loop in $\calm$ at $\slam$., $\gamma : [0,1] 
\rightarrow \calm, \gamma(0) = \gamma(1)$. For this $\gamma$ a holonomy 
operator $\Gamma_{\cala}$ is defined :
\begin{equation}
  \label{eq:holonomy}
  \Gamma_{\cala}(\gamma) = {\cal P}exp\left\{\oint_{\gamma}\cala\right\} 
  \in \kansu{U}{m},
\end{equation}
where ${\cal P}$ means path-ordered. This acts on the fiber $F_0$ at 
$\slam$ of the vector bundle 
$\left\{\fukuso^m,\widetilde{E}, \pi_{\widetilde{E}}, M\right\}$ as follows :
${\textbf x} \rightarrow \Gamma_{\cala}(\gamma){\textbf x}$.\quad 
The holonomy group $Hol(\cala)$ is in general subgroup of $\kansu{U}{m}$ 
. In the case of $Hol(\cala) = \kansu{U}{m}$,   $\cala$ is called 
irreducible, see \cite{MN}.

\par \noindent
In the Holonomic Quantum Computation we take  
\begin{eqnarray}
  \label{eq:information}
  &&{\rm Encoding\ of\ Information} \Longrightarrow {\textbf x} \in F_0 , 
  \nonumber \\
  &&{\rm Processing\ of\ Information} \Longrightarrow \Gamma_{\cala}(\gamma) : 
  {\textbf x} \rightarrow \Gamma_{\cala}(\gamma){\textbf x}.
\end{eqnarray}
%%%%%%%%%%%%%%%%%%%%%%%%%%%%%%%%%%%%%%%%%%%%%%%%%%%%%%%%%%%%5
\begin{center}
\setlength{\unitlength}{1mm}
\begin{picture}(160,130)
\put(80,100){\circle*{4}}
\put(80,60){\circle*{4}}
\put(80,20){\circle*{4}}
\put(80,102){\line(0,1){15}}
\put(80,62){\line(0,1){56}}
\put(80,32){\line(0,1){26}}
\put(79.5,24){$\cdot$}
\put(79.5,26){$\cdot$}
\put(79.5,28){$\cdot$}
\qbezier(80,60)(140,80)(80,100)
\qbezier(80,20)(100,5)(110,20)
\qbezier(80,20)(100,35)(110,20)
\put(72,59){{\bf X}}
\put(69,99){A{\bf X}}
\put(72,18){$\lambda_{0}$}
\put(75,122){$E(\lambda_{0})$}
\qbezier(50,0)(90,10)(150,0)
\qbezier(50,0)(20,20)(30,30)
\qbezier(150,0)(140,20)(120,30)
\qbezier(30,30)(80,40)(120,30)
\put(120,10){${\cal M}$}
\end{picture}
\end{center}
\begin{center}
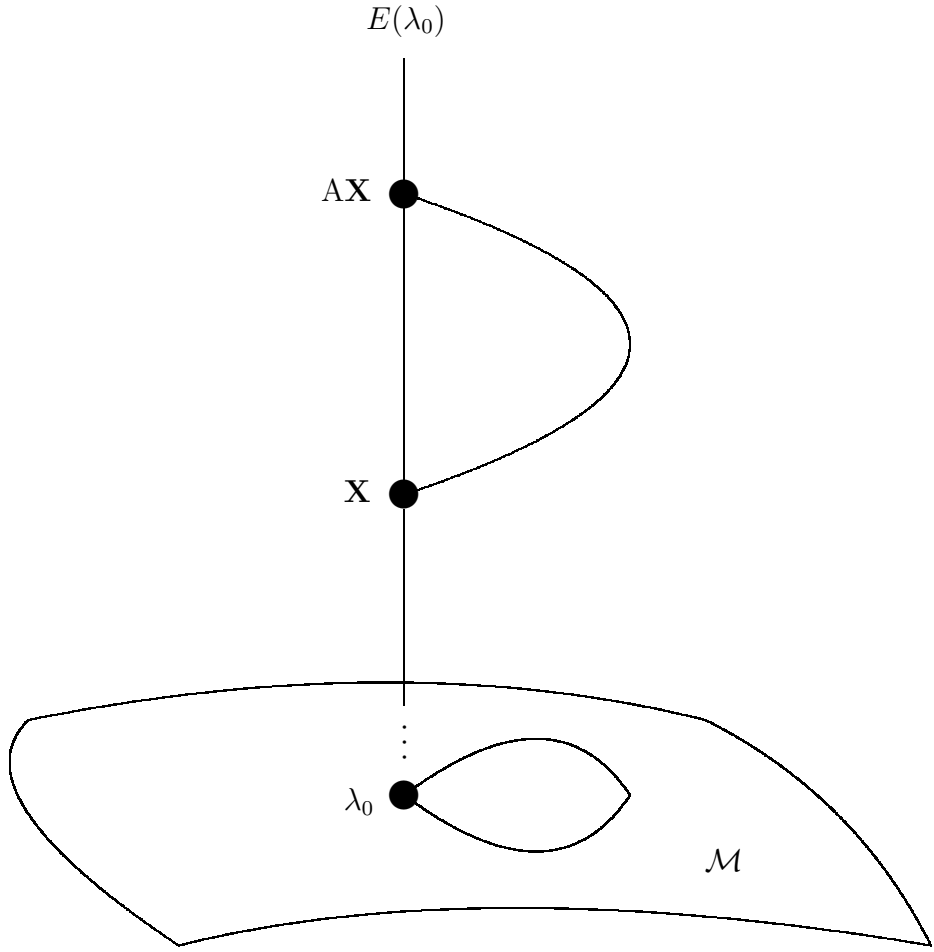
 \textbf{Quantum Computational Bundle}
\end{center}
%%%%%%%%%%%%%%%%%%%%%%%%%%%%%%%%%%%%%%%%%%%%%%%%%%%%%%%%%%
\section{Holonomic Quantum Computation}

We apply the results of last section to Quantum Optics and discuss  
(optical) Holonomic Quantum Computation proposed by \cite{ZR} and 
\cite{PC}. 
Let $a(a^\dagger)$ be the annihilation (creation) operator of the 
harmonic oscillator.
If we set $N\equiv a^\dagger a$ (:\ number operator), then
\begin{equation}
  [N,a^\dagger]=a^\dagger\ ,\
  [N,a]=-a\ ,\
  [a,a^\dagger]=1\ .
\end{equation}
Let $\calh$ be a Fock space generated by $a$ and $a^\dagger$, and
$\{\ket{n}\vert n\in\futon\cup\{0\}\}$ be its basis.
The actions of $a$ and $a^\dagger$ on $\calh$ are given by
\begin{equation}
  \label{eq:shoukou}
  a\ket{n} = \sqrt{n}\ket{n-1}\ ,\
  a^\dagger\ket{n} = \sqrt{n+1}\ket{n+1}\ ,
\end{equation}
where $\ket{0}$ is a vacuum ($a\ket{0}=0$). In the following we treat
coherent operators and squeezed operators.
\begin{eqnarray}
  \label{eq:coherent-operator} 
  &&\mbox{Coherent Operator}\ \ 
  D(\alpha)=\mbox{exp}\left(\alpha a^\dagger - \bar{\alpha}a\right) \quad 
  {\rm for}\  \alpha \in \fukuso , \\
  \label{eq:squeezed-operator}
  &&\mbox{Squeezed  Operator}\ \   
   S(\beta) = \mbox{exp}\left(\beta \frac{1}{2}(a^\dagger)^2 
   - \bar{\beta} \frac{1}{2}a^2 \right)\quad  {\rm for}\  \beta \in \fukuso.
\end{eqnarray}
For the details see \cite{KF1}. 
Next we consider the system of $n$--harmonic oscillators. If we set
\begin{equation}
  \label{eq:n-system}
  a_i = 1\otimes \cdots \otimes 1\otimes a\otimes 1\otimes \cdots \otimes 1, 
        \quad  
  {a_i}^{\dagger} = 1\otimes \cdots \otimes 1\otimes a^{\dagger}\otimes 
                    1\otimes \cdots \otimes 1, 
\end{equation}
for $1 \leq i \leq n$, then it is easy to see 
\begin{equation}
  \label{eq:relations}
 [a_i, a_j] = [{a_i}^{\dagger}, {a_j}^{\dagger}] = 0,\ 
 [a_i, {a_j}^{\dagger}] = \delta_{ij}.
\end{equation}
We also denote by $N_{i} = {a_i}^{\dagger}a_i\ (1 \leq i \leq n)$ 
number operators.
\subsection{Two--Qubit Case}

Since we want to consider coherent states based on Lie algebras $su(2)$ 
and $su(1,1)$, we make use of Schwinger's boson method, see \cite{FKSF1}  
and \cite{FKSF2}. Namely if we set 
\begin{eqnarray}
  \label{eq:J-daisu}
    &&\ su(2) :\quad
     J_+ = a_1^{\dagger}a_2,\ J_- = a_2^{\dagger}a_1,\ 
     J_3 = {1\over2}\left(a_1^{\dagger}a_1 - a_2^{\dagger}a_2\right), \\
  \label{eq:K-daisu}
    &&\ su(1,1) :\quad
     K_+ = a_1^{\dagger}a_2^{\dagger},\ K_- = a_2 a_1,\ 
     K_3 = {1\over2}\left(a_1^{\dagger}a_1 + a_2^{\dagger}a_2  + 1\right),
\end{eqnarray}
then we have
\begin{eqnarray}
  \label{eq:J-relation}
     &&\ su(2) :\quad
     [J_3, J_+] = J_+,\ [J_3, J_-] = - J_-,\ [J_+, J_-] = 2J_3, \\
  \label{eq:K-relation}
     &&\ su(1,1) :\quad
     [K_3, K_+] = K_+,\ [K_3, K_-] = - K_-,\ [K_+, K_-] = -2K_3.
\end{eqnarray}

In the following we treat unitary coherent operators based on Lie algebras 
$su(2)$ and $su(1,1)$. 
\begin{eqnarray}
  \label{eq:J-operator}
   U(\lambda) &=& \mbox{exp}\left(\lambda J_{+} - \bar{\lambda}J_{-}\right)
   \quad \mbox{for}\  \lambda \in \fukuso , \\
  \label{eq:K-operator}
   V(\mu) &=& \mbox{exp}\left(\mu K_{+} - \bar{\mu}K_{-}\right)
    \quad \mbox{for}\  \mu \in \fukuso.
\end{eqnarray}
For the details of $U(\lambda)$ and $ V(\mu)$ see \cite{AP} and \cite{FKSF1}. 

Let $H_0$ be a Hamiltonian with nonlinear interaction produced by 
a Kerr medium., that is  $H_0 = \hbar {\rm X} N(N-1)$, where X is a 
certain constant, see \cite{MW} and \cite{PC}. The eigenvectors of $H_0$ 
corresponding to $0$ is $\left\{\ket{0},\ket{1}\right\}$, so its eigenspace is 
${\rm Vect}\left\{\ket{0},\ket{1}\right\} \cong \fukuso^2$. 
The space ${\rm Vect}\left\{\ket{0},\ket{1}\right\}$ is called 
1-qubit (quantum bit) space, see \cite{AS} or \cite{RP}. 
Since we are considering the system of two particles, the Hamiltonian that 
we treat in the following is 
\begin{equation}
  \label{eq:hamiltonian}
  H_0 = \hbar {\rm X} N_{1}(N_{1}-1) + \hbar {\rm X} N_{2}(N_{2}-1).
\end{equation}
The eigenspace of $0$ of this Hamiltonian becomes therefore
\begin{equation}
  \label{eq:2-eigenspace}
   F_0 = {\rm Vect}\left\{\ket{0},\ket{1}\right\}\otimes 
         {\rm Vect}\left\{\ket{0},\ket{1}\right\} 
    = {\rm Vect}\left\{\kett{0}{0},\kett{0}{1}, \kett{1}{0},
                       \kett{1}{1}\right\}
    \cong \fukuso^{4}. 
\end{equation}
We set $\ket{vac} = 
\left(\kett{0}{0}, \kett{0}{1}, \kett{1}{0}, \kett{1}{1} \right)$.
Next we consider the following isospectral family of $H_0$ : 
\begin{eqnarray}
  \label{eq:twofamily}
   &&H_{(\alpha_{1},\beta_{1},\lambda,\mu,\alpha_{2},\beta_{2})}
  = W(\alpha_{1},\beta_{1},\lambda,\mu,\alpha_{2},\beta_{2})H_0 
     W(\alpha_{1},\beta_{1},\lambda,\mu,\alpha_{2},\beta_{2})^{-1},\\
  \label{eq:double}
  &&W(\alpha_{1},\beta_{1},\lambda,\mu,\alpha_{2},\beta_{2}) 
  = W_{1}(\alpha_{1},\beta_{1})O_{12}(\lambda,\mu)
      W_{2}(\alpha_{2},\beta_{2}). 
\end{eqnarray}
where 
\begin{equation}
   O_{12}(\lambda, \mu) = U(\lambda)V(\mu), \quad 
   W_{j}(\alpha_{j},\beta_{j}) = D_{j}(\alpha_{j})S_{j}(\beta_{j}) \quad 
     \mbox{for}\quad j = 1, 2.   
\end{equation}
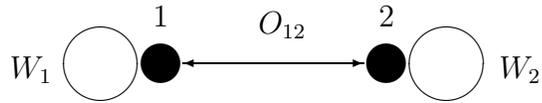
\begin{center}
\setlength{\unitlength}{1mm}
\begin{picture}(80,22)
  \put(30,8){\circle*{6}}
  \put(60,8){\circle*{6}}
  \put(22,8){\circle{10}}
  \put(68,8){\circle{10}}
  \put(45,8){\vector(-1,0){12}}
  \put(45,8){\vector(1,0){12}}
  \put(29,13){1}
  \put(59,13){2}
  \put(10,6){$W_{1}$}
  \put(75,6){$W_{2}$}
  \put(43,12){$O_{12}$}
\end{picture}
\end{center}
In this case
\begin{equation}
 {\cal M}=\left\{(\alpha_{1},\beta_{1},\lambda,\mu,\alpha_{2},\beta_{2}) 
 \in {\mathbf C}^{6} \right\}
\end{equation}
and we want to calculate 
\begin{equation}
 \label{eq:2-connection}
  {\cal A}=\bra{vac}W^{-1}dW\ket{vac},
\end{equation}
where 
\begin{eqnarray}
    d&=&d\alpha_{1}\frac{\partial}{\partial \alpha_{1}}+
      d\bar{\alpha_{1}}\frac{\partial}{\partial \bar{\alpha_{1}}}+
      d\beta_{1}\frac{\partial}{\partial \beta_{1}}+
      d\bar{\beta_{1}}\frac{\partial}{\partial \bar{\beta_{1}}}+ 
      d\lambda\frac{\partial}{\partial \lambda}+
      d\bar{\lambda}\frac{\partial}{\partial \bar{\lambda}}+
      d\mu\frac{\partial}{\partial \mu}+
      d\bar{\mu}\frac{\partial}{\partial \bar{\mu}}  \nonumber \\
     &+&
      d\alpha_{2}\frac{\partial}{\partial \alpha_{2}}+
      d\bar{\alpha_{2}}\frac{\partial}{\partial \bar{\alpha_{2}}}+
      d\beta_{2}\frac{\partial}{\partial \beta_{2}}+
      d\bar{\beta_{2}}\frac{\partial}{\partial \bar{\beta_{2}}}\ . 
\end{eqnarray}
The calculation of (\ref{eq:2-connection}) is not easy, see \cite{KF5} 
for the details.

\par \noindent
\textbf{Problem}\quad Is the connection form irreducible in $U(4)$ ?
\par \noindent
Our analysis in \cite{KF5} shows that the holonomy group generated 
by ${\cal A}$ may be $SU(4)$ not $U(4)$. To obtain $U(4)$ a sophisticated 
trick $\cdots$ higher dimensional holonomies \cite{AFG} $\cdots$ may be 
necessary. A further study is needed.  
\subsection{N--Qubit Case}

A reference Hamiltonian is in this case 
\[
  H_0 = X \sum_{i=1}^{n} N_i(N_i-1),\quad  X\  \mbox{is\ a\  constant} 
\]
and the eigen--space to 0--eigenvalue ($n$--qubits) becomes 
\begin{equation}
  F_0 = \mbox{Vect}\{\ket{0,\cdots,0,0},\ket{0,\cdots,0,1},\cdots, 
                     \ket{1,\cdots,1,0},\ket{1,\cdots,1,1}\} 
      \cong \fukuso^{2^n}  
\end{equation}
and set $
  \ket{vac} = ( \ket{0,\cdots,0,0},\ket{0,\cdots,0,1},\cdots,
                \ket{1,\cdots,1,0},\ket{1,\cdots,1,1})
$ .

\par \noindent
The $u(n)$--algebra is defined by 
\begin{equation}
  \mbox{generators}\ :\ \{E_{ij}\ |\ 1\leq i, j \leq n \}, \quad  
  \mbox{relations}\ :\ [E_{ij},E_{kl}]=\delta_{jk}E_{il}- \delta_{li}E_{kj},
\end{equation}
and $\{E_{in}\ |\ 1\leq i \leq n-1 \}$ a Weyl basis. 
Boson representation of $u(n)$--algebra is well--known to be 
\begin{equation}
  E_{ij}={a_{i}}^{\dagger}a_{j}\quad  1\leq i, j \leq n \ .   
\end{equation}
The $u(n-1,1)$--algebra is also defined by 
\begin{equation}
   \mbox{generators}\ :\ \{E_{ij}\ |\ 1\leq i, j \leq n \}, \quad  
   \mbox{relations}\ :\ [E_{ij},E_{kl}]=\eta_{jk}E_{il} - \eta_{li}E_{kj}, 
\end{equation}
where $\eta=\mbox{diag}(1,\cdots,1,-1)$, 
and $\{E_{in}\ |\ 1\leq i \leq n-1 \}$ a Weyl basis. 
Boson representation of $u(n-1,1)$--algebra is given by 
\begin{eqnarray}
  &&E_{ij}={a_{i}}^{\dagger}a_{j} \quad  1\leq i, j \leq n-1\ , \quad 
    E_{nn}={a_{n}}^{\dagger}a_{n}+1    \nonumber \\
  &&E_{in}={a_{i}}^{\dagger}{a_{n}}^{\dagger}\ , \quad   
    E_{ni}=a_{n}a_{i} \quad  1\leq i \leq n-1\ . 
\end{eqnarray}

\par \noindent
A family of Hamiltonians that we treat is 
\begin{equation}
     H = WH_{0}W^{-1}\ ,  \qquad W=\prod_{j=1}^{n} W_{jn} 
\end{equation}
and $W_{jn}$ is 
\begin{equation}
  W_{jn}(\alpha_{j},\beta_{j},\lambda_{j},\mu_{j}) 
  =   W_{j}(\alpha_{j},\beta_{j})O_{jn}(\lambda_{j},\mu_{j}),
\end{equation}
and 
\begin{equation}
  W_{j}(\alpha_{j},\beta_{j}) = D_{j}(\alpha_{j})S_{j}(\beta_{j}), \quad 
  O_{jn}(\lambda_{j},\mu_{j}) = U_{j}(\lambda_{j})V_{j}(\mu_{j}),
\end{equation}
and for $1 \leq j \leq n-1$
\begin{equation}
 U_{j}(\lambda_{j}) = \mbox{exp}(\lambda_{j}{a_j}^{\dagger}a_n - 
       \bar{\lambda}_{j}{a_n}^{\dagger}a_j), \quad  
 V_{j}(\mu_{j}) = \mbox{exp}(\mu_{j}{a_j}^{\dagger}{a_n}^{\dagger} - 
     \bar{\mu}_{j}a_na_j) 
\end{equation}
making use of each Weyl basis. We note that $O_{nn}=1$. 
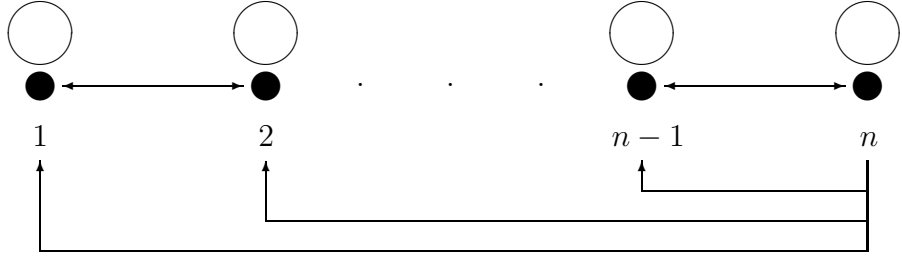
\begin{center}
\setlength{\unitlength}{1mm}
\begin{picture}(130,45)
  \put(10,28){\circle*{4}}
  \put(9,20){$1$}
\put(10,35){\circle{8}}
\put(25,28){\vector(-1,0){12}}
\put(25,28){\vector(1,0){12}}
  \put(40,28){\circle*{4}}
  \put(39,20){$2$}
\put(40,35){\circle{8}}
  \put(52,28){.}
  \put(64,28){.}
  \put(76,28){.}
  \put(90,28){\circle*{4}}
  \put(90,35){\circle{8}}
\put(105,28){\vector(1,0){12}}
\put(105,28){\vector(-1,0){12}}
  \put(86,20){$n-1$}
  \put(120,28){\circle*{4}}
  \put(119,20){$n$}
  \put(120,18){\line(0,-1){12}}
  \put(120,6){\line(-1,0){110}}
  \put(10,6){\vector(0,1){12}}
  \put(120,10){\line(-1,0){80}}
  \put(40,10){\vector(0,1){8}}
  \put(120,14){\line(-1,0){30}}
  \put(90,14){\vector(0,1){4}}
\put(120,35){\circle{8}}
\end{picture}
\end{center}
In this case
\begin{equation}
  {\cal M}=
   \left\{(\cdots,\alpha_{j},\beta_{j},\lambda_{j},\mu_{j},
  \alpha_{j+1},\beta_{j+1}, \cdots)\right\}
  \cong {\mathbf C}^{4n-2} \nonumber 
\end{equation}
and
\begin{equation}
  {\cal A}=\bra{vac}W^{-1}dW\ket{vac}\ , 
\end{equation}
where
\begin{eqnarray}
    d&=& \sum_{j=1}^{n}\
     d\alpha_{j}\frac{\partial}{\partial \alpha_{j}}+
      d\bar{\alpha}_{j}\frac{\partial}{\partial \bar{\alpha}_{j}}+
      d\beta_{j}\frac{\partial}{\partial \beta_{j}}+
      d\bar{\beta}_{j}\frac{\partial}{\partial \bar{\beta}_{j}} \nonumber \\
    &+&\sum_{j=1}^{n-1}\ 
      d\lambda_{j}\frac{\partial}{\partial \lambda_{j}}+
      d\bar{\lambda}_{j}\frac{\partial}{\partial \bar{\lambda}_{j}}+
      d\mu_{j}\frac{\partial}{\partial \mu_{j}}+
      d\bar{\mu}_{j}\frac{\partial}{\partial \bar{\mu}_{j}}\ . 
\end{eqnarray}
\par \vspace{3mm} \noindent
We propose the following 
\par \noindent
\textbf{Problem}\quad Is the connection form irreducible in $U(2^n)$ ?
\par \noindent
%
%
%%%%%%%%%%%%
%References%
%%%%%%%%%%%%


\begin{thebibliography}{99}
\bibitem{PS}P. W. Shor : %1
\newblock Polynomial-time algorithms for prime factorization and discrete 
logarithms on a quantum computer,
\newblock SIAM J. Computing., 26(1997), 1484,
\newblock quant-ph/9508027.
%
\bibitem{AS} A. Steane : %2
\newblock Quantum Computing,
\newblock Rept. Prog. Phys., 61(1998), 117,  
\newblock quant-ph/9708022. 
%
\bibitem{RP} E. Rieffel and W. Polak : %3
\newblock An Introduction to Quantum Computing for Non-Physicists,
\newblock quant-ph/9809016.
%
\bibitem{KF9}K. Fujii : %1
\newblock Introduction to Grassmann Manifolds and Quantum Computation, 
\newblock quant-ph/0103011.
%
\bibitem{ZR}P. Zanardi and M. Rasetti : %6
\newblock Holonomic Quantum Computation,
\newblock Phys. Lett. A264(1999), 94,
\newblock quant-ph/9904011.
%
\bibitem{PZR}J. Pachos, P. Zanardi and M. Rasetti : %7
\newblock Non-Abelian Berry connections for quantum computation,
\newblock Phys. Rev. A 61(2000), 010305(R),  
\newblock quant-ph/9907103.
%
\bibitem{AYK}A. Yu. Kitaev : %9
\newblock Fault-tolerant quantum computation by anyons,
\newblock quant-ph/9707021. 
%
\bibitem{JP}J. Preskill : %8
\newblock Fault-Tolerant Quantum Computation,
\newblock quant-ph/97120408. 
%
\bibitem{SW}A. Shapere and F. Wilczek (Eds) : %10 
\newblock Geometric Phases in Physics,
\newblock World Scientific, Singapore, 1989.
%
\bibitem{MN}M. Nakahara : %14
\newblock Geometry, Topology and Physics,
\newblock IOP Publishing Ltd, Bristol and New York, 1990.
%
\bibitem{MW}L. Mandel and E. Wolf : 
\newblock Optical Coherence and Quantum Optics, 
\newblock Cambridge University Press, 1995.
%
\bibitem{PC}J. Pachos and S. Chountasis : %11
\newblock Optical Holonomic Quantum Computer,
\newblock Phys. Rev. A 62(2000), 052318, 
\newblock quant-ph/9912093.
%
\bibitem{KF1} K. Fujii : %12
\newblock Note on Coherent States and Adiabatic Connections, Curvatures,
\newblock J. Math. Phys.,  
\newblock 41(2000), 4406, 
\newblock quant-ph/9910069.
%
\bibitem{KF3} K. Fujii : %13
\newblock Mathematical Foundations of Holonomic Quantum Computer,
\newblock to appear in Rept. Math. Phys,
\newblock quant-ph/0004102.
%
\bibitem{KF4} K. Fujii : %13
\newblock More on Optical Holonomic Quantum Computer,
\newblock quant-ph/0005129.
%
\bibitem{KF5} K. Fujii : %13
\newblock Mathematical Foundations of Holonomic Quantum Computer II,
\newblock quant-ph/00101102.
%
\bibitem{AP}A. Perelomov : %18
\newblock Generalized Coherent States and Their Applications,
\newblock Springer--Verlag, 1986.
%
\bibitem{FKSF1}K. Funahashi, T. Kashiwa, S. Sakoda and K. Fujii : %16
\newblock Coherent states, path integral, and semiclassical approximation,
\newblock  J. Math. Phys., 36(1995), 3232.
%
\bibitem{FKSF2}K. Funahashi, T. Kashiwa, S. Sakoda and K. Fujii : %17
\newblock Exactness in the Wentzel-Kramers-Brillouin approximation for 
some homogeneous spaces,
\newblock J. Math. Phys., 36(1995), 4590.
%
\bibitem{FKS}K. Fujii, T. Kashiwa, S. Sakoda :%16
\newblock Coherent states over Grassmann manifolds and the WKB exactness
in path integral,
\newblock J. Math. Phys., 37(1996), 567.
%
\bibitem{PZ}J. Pachos and P. Zanardi : %7
\newblock Quantum Holonomies for Quantum Computing,
\newblock quant-ph/0007110. 
%
\bibitem{AFG}O. Alvarez, L. A. Ferreira and J. S. Guillen : 
\newblock A New Approach to Integrable Theories in Any Dimension,
\newblock Nucl. Phys. B529(1998), 689, 
\newblock hep-th/9710147.
%
\end{thebibliography}
\end{document}